\documentclass[prb, preprint]{revtex4}
\usepackage{amsfonts}
\usepackage{amsmath}
\usepackage{amssymb}
\usepackage{graphicx}
\usepackage{epstopdf}
\usepackage{CJK}
\usepackage{bm}
\usepackage{bm,color}
\begin{document}
\begin{CJK*}{GBK}{}

\title{Magnetic phases and unusual topological electronic structures of  Weyl semimetals in strong interaction limit}
\author{Liang-Jun Zhai$^{1,2}$, Po-Hao Chou$^{1}$, and Chung-Yu Mou$^{1,3,4}$%
}
\affiliation{$^{1}$Department of Physics, National Tsing Hua University, Hsinchu 30043,
Taiwan, 300, R.O.C.}
\affiliation{$^{2}$The School of Mathematics and Physics, Jiangsu University of Technology,
Changzhou 213001, China}
\affiliation{$^{3}$Institute of Physics, Academia Sinica, Nankang, Taiwan, R.O.C.}
\affiliation{$^{4}$Physics Division, National Center for Theoretical Sciences, P.O.Box
2-131, Hsinchu, Taiwan, R.O.C.}

\begin{abstract}
The interplay of electronic band structures and electron-electron interactions is known to
brew new phases in condensed matter. In this paper, we investigate thermodynamic phases
and corresponding electronic structures of the Weyl semimetal in
the strong onsite Coulomb interaction limit.
Based on a minimum model of the Weyl semimetal with two linear Weyl nodes,
it is shown that generically the Weyl semimetal becomes magnetic in the presence
of interactions. In particular, it is shown that the Dzyaloshinskii-Moriya exchange interaction is generally induced
so that the A-type antiferromagnetic (A-AFM) phase and the spiral spin density wave (SSDW) states are two generic phases.
Furthermore, we find that Weyl nodes proliferate and it is possible to doubly enhance the
unusual properties of non-interacting Weyl semimetals through the realization of double-Weyl nodes in strong correlation limit. Specifically, it is shown that in the SSDW phase, linear Weyl nodes are tuned into double-Weyl nodes with the corresponding charges being $\pm 2$.
As the spin-orbit coupling increases, a quantum phase transition occurs with the SSDW phase being
turned into an A-AFM phase and at the same time, double-Weyl nodes are disintegrated into
two pairs of linear Weyl nodes. Our results reveal
the unusual interplay between the topology of electronic structures and magnetism in strongly correlated phases of Weyl semimetals.
\end{abstract}

\pacs{74.70.Xa, 74.20.Mn, 74.20.Rp}
\maketitle

\section{Introduction}
Since the discovery of
the time reversal (TR) invariant topological insulator (TI)
in two dimensions (2D) and three dimensions (3D) \cite{Hasan,Hasan1}, the topological aspects of the electronic band structures have become important benchmarks to characterize electronic
phases in condensed matter physics. The unique feature of TIs lies in their properties of being insulating in the bulk, and yet their surfaces being metallic due to the existence of surface states. The surface states
result from the nontrivial topology of the bulk band structure and are robust against
perturbations that respect symmetries of the system.
More recently, it is further realized that even if the band structure of the materials are gapless,
non-trivial topology of the gapless points (nodal points) may also result in topologically protected surface states\cite{Murakami}.

Among the class of topological gapless materials, the Dirac semimetal and the Weyl semimetal are the simplest type, characterized by the presence of nodal points at which two distinct bands touch each\cite{Burkov,Wan,Qi,Wang,Castro Neto,Mou}.
For the Dirac semimetal, the low-energy Hamiltonian near isolated Dirac nodes
can be described by the Dirac equation with both of time-reversal symmetry and inversion symmetry
being preserved and it can be realized in both the 2D and 3D systems, such as graphene (2D) \cite{Castro Neto}
and $\rm Na_3Bi$ (3D) \cite{Liu}.
In contrast, the time-reversal symmetry or inversion symmetry is explicitly broken in the
Weyl semimetal, and the Hamiltonian around the Weyl nodes
is described by the Weyl Hamiltonian
given by $H=\pm v_F\bm{\sigma}\mathbf{\cdot k}$ with $\bm{\sigma}$ being the Pauli matrices
and $\textbf{k}$ being the momentum deviation from the Weyl point, and $\pm$ denotes the chirality.
The absence of the time-reversal symmetry or the inversion symmetry results in the separation of the Weyl
nodes either in momentum or in energy. In addition, it results in the non-trivial topology (chirality) carried by
the Weyl nodes, which is characterized by the monopole charges $Q$ defined as the integral of the Berry curvature $\bm{\Omega(k)}$ over the surface enclosing the node, $Q=(1/2\pi)\oint d \bm{S_k\cdot \Omega(k)}$ \cite{Xiao}.
 Due to the topological nature of $Q$, Weyl fermions in these materials are robust to small perturbations\cite{Kobayashi, Goswami}.
For large
perturbations, it is known that localized states may arise near point defects\cite{Huang} and
Weyl points can even appear or disappear in pairs with opposite monopole charges.
In the simple Weyl semimetals, each Weyl point carries $Q=1$ or $Q=-1$, which
has been first realized experimentally in $\rm TaAs$ \cite{Yang}.
The possibility of the multi-Weyl nodes with $|Q|>1$
has also been proposed \cite{Fang, Xu, Jian} recently. Instead of being the usual linear Weyl nodes carrying
$\pm 1$ monopole charge, the multi-Weyl nodes, protected by $C_4$ or $C_6$ rotation symmetry, have
nonlinear dispersion and higher monopole charge. The lowest non-trivial example is
the double-Weyl semimetal with $Q=\pm2$, which is suggested to be realized
in the 3D semimetal $\rm{HgCr_2Se_4}$ \cite{Xu}. The double-Weyl nodes possess quadratic dispersions
in two direction, e.g., the $xy$ plane and linear dispersion in the third direction. Furthermore,
it is known that due to larger chiral charges $Q$ in presence, the unusual phenomena (
such as the quantum anomalous Hall conductivity,
the chiral anomaly, and the number of Fermi arcs) associated with the linear Weyl nodes
are doubly enhanced \cite{Jian, Chang2, Lai}.

While the above mentioned properties are valid for noninteracting Weyl semimetals, it is known that the non-trivial topology in electronic structure is driven by the spin-orbit interaction, which
inevitably involves heavy elements. The correlation effects due to interactions are
therefore present and it is necessary to examine effects of interactions
on properties of the Weyl semimetals.
For topological insulators, the Coulomb interaction is usually screened and
becomes short-ranged. Studies indicate that topological transitions
may be induced so that nontrivial
topological phases may be broken into topologically trivial phases
by sufficiently strong short-range correlation\cite{Mou, Yu, Yamaji,Yoshida, Huang1}.
On the other hand, for the Dirac/Weyl semimetals, the long-range Coulomb interaction is
shown to be marginally irrelevant and induces logarithmic corrections in response functions\cite{Mou, Sheehy, Jian,Lai}.
In the strong coupling limit, however, the semimetals could be turned into either
a Mott insulator if the nodal point is anisotropic \cite{Sekine} or a charge density wave (CDW) state\cite{Wei}.
Since in typical materials that realize Dirac/Weyl semimetals, localized electronic orbits, such as $d$ orbits in $Ta$ of $TaAs$, are often involved, the onsite Coulomb energy in the materials usually dominates. Therefore, it would
be interesting and necessary to investigate effects of short-ranged Coulomb
interaction on the Weyl semimetal.

In this paper, we examine phases of the Weyl semimetal in the presence
of the onsite Coulomb interaction in a Hubbard model. Similar to effects of disorders\cite{Kobayashi}, in the weak interacting regime, it is found that 
the Weyl semimetal is paramagnetic without magnetic orders.
The electronic structure is similar to that of the non-interacting Weyl fermions with parameters be renormalized. As the on-site interaction increases and is strong
enough, the Weyl semimetal becomes magnetic.
In the strong interaction
limit of a Hubbard model, we find that the A-AFM phase and
the SSDW phase are two generic phases. In the SSDW phase,
each Weyl node is tuned into a double-Weyl node with the corresponding charge being
$\pm 2$. As the spin-orbit coupling increases, a first-order phase transition occurs
with the SSDW phase being turned into an A-AFM phase and at the same
time, double-Weyl nodes are disintegrated into
two pairs of linear Weyl nodes. Our results reveal
the unusual interplay between the topology of electronic structures and magnetism in strongly correlated
phases of Weyl semimetals and pave a way for realizing the double-Weyl semimetal
based on a linear Weyl semimetals.

The rest of the paper is organized as follows. In Sec. \ref{Sec:II}, the model of the Weyl semimetal with a Hubbard interaction is introduced. By using a canonical transformation, we derive the effective low-energy Hamiltonian for the strong coupling limit of the Hubbard model.  Under the Gutzwiller approximation,
the renormalized mean-field Hamiltonian is constructed. In Sec. III, magnetic phases in the strong Coulomb interaction limit are solved numerically. It is shown that the corresponding electronic structures possess non-trivial nodal structures.
Finally, in Sec. IV, we conclude and discuss. How our results change from the weak interaction regime to the strong coupling regime is presented and discussed. Possible connection of our results to experimental observations is also presented.

\section{Theoretical Model of Weyl semimetal}
\label{Sec:II}
We start with a minimum model of 3D Weyl semimetal on a simple cubic lattice\cite{Chang} with an onsite Hubbard
repulsion interaction $U$.  The model has the minimum number of two Weyl points. Since the
onsite Hubbard $U$ interaction is rotationally invariant, two Weyl points can be chosen to be along $z$ axis so that
the Hamiltonian is given by
\begin{eqnarray}
  H =\sum_{{\bf k}, \alpha, \beta} C^{\dagger}_{\alpha} ({\bf k}) H_{0, \alpha \beta} ({\bf k}) C_{\beta} ({\bf k}) +U\sum_{i}{\hat n_{i\uparrow}\hat n_{i\downarrow}}  \label{model}
 \end{eqnarray}
with $H_0$ being a $2 \times 2$ matrix
\begin{eqnarray}
  H_0({\bf k})= (t_1\cos k_z - \mu) \sigma_0
  +t_2(m+2-\cos{k_x}-\cos{k_y})\sigma_z
  +t_{so}\sin{{\bf k}}\cdot {\bm \sigma}. \label{model_2}
  \end{eqnarray}
Here $\alpha$ and $\beta$ takes $\uparrow$ or $\downarrow$ that represent the spin up or down, and $C_{\alpha}^\dagger$ creates an electron with spin up or down. 
$\sin{{\bf k}}\cdot {\bm \sigma}$ is a shorthand notation for $\sigma_x \sin k_x+\sigma_y\sin k_y+\sigma_z \sin k_z$. $t_1$ and $t_2$ represent the hopping amplitudes along the $z$ direction and in the $xy$ plane respectively.
$t_{so}$ is the strength of the spin-orbit coupling and $m$ is the exchange parameter that controls the degree of time-reversal symmetry breaking. $m$ will be set to zero in most of the following computations.
$\mu$ is the chemical potential, and $\sigma_0$ is the $2 \times 2$ identity matrix.
Finally, $U$ describes the onsite Hubbard repulsion interaction.
The model, Eq.(\ref{model_2}), is an extension of the Qi-Wu-Zhang (QWZ) model in the study of the 2D
quantum anomalous Hall effect \cite{Qi2} and it has only two Weyl nodes at the ground state with $U=0$ and $m=0$ \cite{Chang}.

In the case of $U=0$, the system has a $C_4$ rotation symmetry, and the nonzero $t_1$
would break the space inversion
symmetry with respect to the $x-y$ plane, while the $t_2$ term breaks the time-reversal symmetry.
The energies of the two bands can be solved and are given by
\begin{eqnarray}
E_{\pm}(\bf {k})&=&t_1\cos{k_z}\pm\sqrt{t_{so}^2(\sin^2{k_x}+\sin^2{k_y})
+[t_{so}\sin{k_z}+t_2(m+2-\cos{k_x}-\cos{k_y})]^2}. \nonumber \\
\label{energy}
\end{eqnarray}
The ground sate of this minimum model has been well-studied in Ref.[\onlinecite{Chang}].
By tuning values of $m$ and $t_{so}$, the ground state of the two-band model could be
either in a topological trivial phase or the Weyl semimetal phase.

When $U$ is switched on and is small, the free energy gain due to magnetic orders is 
also small. One thus expects that the Weyl semimetal is paramagnetic without magnetic orders.
Hence the electronic structure is similar to that of the non-interacting Weyl fermions with parameters be just renormalized. As $U$ increases, the free energy gain due to magnetic orders also increases. As a result, only when $U$ is strong enough, the Weyl semimetal starts to become magnetic. Therefore, we shall first consider the strong interaction limit when $U$ is large. The connection of strong $U$ limit to weak $U$ limit will be discussed later.

In the strong interaction limit when $U$ is large, the band structure resulted from Eq. (\ref{model})
can be very different from that for $U=0$. In the large $U$ limit, the Hilbert space is energetically decomposed into singly occupied and doubly occupied spaces so that the electron operator can be decomposed as
$C^{\dagger}_{i\sigma} = C^{\dagger}_{i\sigma} (1- n_{i,-\sigma}) + C^{\dagger}_{i\sigma} n_{i,-\sigma}$.
Since only the kinetic energy, $T \equiv \sum_{i,j} C^{\dagger}_{i} H_{0,ij}  C_j$, mixes singly occupied and doubly occupied spaces, we first perform a canonical
transformation on the Hamiltonian $H$ to eliminate the mixing term.
If the canonical transformation is generated by $S$, the transformed Hamiltonian $H'$ is given by
\begin{eqnarray}
H'=e^{iS}He^{-iS}=H+[iS,H]+[iS,[iS,H]]+ \cdots . \label{canonical}
\end{eqnarray}
The mixing terms in the kinetic energy can be written as a summation of $T_{+1}$ and $T_{-1}$
with
\begin{eqnarray}
T_{+1} & = &  \sum_{i,j,\alpha,\beta} C^{\dagger}_{i,\alpha} n_{i,-\alpha} H_{0,ij, \alpha, \beta}  C_{j,\beta} (1-n_{i,-\beta}),   \nonumber  \\
T_{-1} & = &  \sum_{i,j,\alpha,\beta} C^{\dagger}_{i,\alpha} (1-n_{i,-\alpha}) H_{0,ij, \alpha, \beta}  C_{j,\beta} n_{i,-\beta}.
\end{eqnarray}
By requiring $T_{+1}  + T_{-1} + [iS,H_U]=0$, one can eliminate the mixing term to first order in $U$. Here $H_U =U\sum_{i}{\hat n_{i\uparrow}\hat n_{i\downarrow}} $ and we find
\begin{eqnarray}
iS = \frac{1}{U} (T_{+1}  - T_{-1}). \label{S}
\end{eqnarray}
Substituting $iS$ in Eq.(\ref{S}) back to Eq.(\ref{canonical}) and keeping the lowest order terms, after some algebra, we find that the low energy Hamiltonian is an extended $t-J$ model which can be
generally expressed as $ H_{eff} = H_t+H_J$ with\cite{Kao} $H_t$ and $H_J$ being given by
\begin{eqnarray}
  H_{t} &=& \sum_{i\delta_z,\sigma}{t_{1}\tilde{C}_{i,\sigma}^+\tilde{C}_{i+\delta_{z}\sigma}}
  +\sum_{i\delta_x(y),\sigma}{\sigma t_2\tilde{C}_{i,\sigma}^+\tilde{C}_{i+\delta_{x(y)}\sigma}}
  -\frac{i}{2}t_{so}\sum_{ij,\sigma\sigma'}{\sigma\tilde{C}_{i,\sigma}^+\tilde{C}_{j \overline{\sigma'}}}
  +\textrm{H.c.},
\end{eqnarray}
and
\begin{eqnarray}
\label{Eq:magneticHamiltonian}
  H_J  &=&\sum_{i,\delta_{x(y)}}{\left( -J_b S_{i}^xS_{i+\delta_{x(y)}}^x
  -J_a S_{i}^yS_{i+\delta_{x(y)}}^y
  +J_b S_{i}^zS_{i+\delta_{x(y)}}^z-\frac{1}{4} J_1 n_in_{i+\delta_{x(y)}}\right)}\\ \nonumber
  &&+\sum_{i,\delta_{z}}{\left[J_z^1 (S_{i}^xS_{i+\delta_{z}}^x+S_{i}^yS_{i+\delta_{z}}^y)
  +J_z^2(S_{i}^z S_{i+\delta_{z}}^z-\frac{1}{4}n_in_{i+\delta_z})+ J_2\delta_z\cdot S_i\times S_{i+\delta_z}\right]}.
\end{eqnarray}
Here $H_t$ and $H_J$ represent the hopping and exchange magnetic interactions of the Hamiltonian.
$S_i^\alpha=\sum_{\sigma\sigma'} \tilde{C}_{i\sigma}^+\sigma_\alpha \tilde{C}_{i\sigma'} (\alpha=x,y,z)$ is the spin operator on site $i$, and $\delta_{\alpha}=(r_j-r_i)_\alpha,(\alpha=x,y,z)$ represents the vector connecting sites in nearest neighbors with $r_i$ being the position of lattice site $i$.
$\tilde{C}_{i\sigma}=(1-n_{i,-\sigma})C_{i\sigma}$ satisfies the no-double-occupancy constraint
for electrons.
In terms of the onsite Hubbard $U$, the strengths of exchange magnetic interactions are given by
$J_a=(t_2^2+t_{so}^2)/U$, $J_b=(t_2^2-t_{so}^2)/U$, $J_1=t_2^2/U$, $J_z^1=(t_1^2-t_{so}^2)/U$,
$J_z^2=(t_1^2+t_{so}^2)/U$, and $J_2=t_{so}^2/U$ with $J_2$ being the Dzyaloshinskii-Moriya (DM) interaction
and the rest of exchanges being the summation of the Heisenberg interaction and spin dipole-dipole interaction.

From Eq.(\ref{Eq:magneticHamiltonian}), it is clear that the DM interaction only appears along the $z$ direction.
Since the DM interaction tends to induce the magnetic spiral phase and the value of $J_z^2$ is
always larger than $J_z^1$, there should be competition between the spiral phase and
the (AFM/FM phase along the $z$ direction.
On the other hand, in the $xy$ plane, the magnetic phase should be AFM or FM,  since there is only Heisenberg
interactions between nearest neighbor sites.
The value of $J_a$ is always larger than $J_b$, therefore, FM holds advantage in the $xy$ plane.

To satisfy the no-double-occupancy constraint, Gutzwiller approximations are adopted
by using the renormalized parameters\cite{Edegger}.
In this approximation, the strength of the spin interaction
remains the same as in the low-doping regime, and the operator $\tilde{C}_{i\sigma}$
is replaced by $C_{i\sigma}$.
Therefore, the hopping Hamiltonian becomes,
\begin{eqnarray}
  H'_t &=& \sum_{i\delta_z,\alpha}{t'_{1}{C}_{i,\alpha}^+{C}_{i+\delta_{z}\alpha}}
  +\sum_{i\delta_x(y),\alpha}{\alpha t'_2{C}_{i,\alpha}^+{C}_{i+\delta_{x(y)}\alpha}}
  -\frac{i}{2}t'_{so}\sum_{ij,\alpha\beta}{\sigma{C}_{i,\alpha}^+{C}_{i \beta}}
  +\textrm{H.c.},
\end{eqnarray}
where $t'_1=g_tt_1$, $t'_2=g_tt_2$ and $t'_{so}=g_tt_{so}$ with $g_t=\frac{1-n}{1-2n_{i\uparrow}n_{i\downarrow}/n}$,
and $n$ is the number of particle on site $i$ (i.e. density of particle number), $n_{i\uparrow}$ and $n_{i\downarrow}$
are numbers of particles for spin up and down electrons respectively.

We shall compute the magnetic phase and electronic structures in a mean approximation with
\begin{eqnarray}
\label{Eq:mean-field}
S_i^\alpha S_j^\beta &\approx&  S_{i}^\alpha\langle S_j^\beta\rangle+S_{j}^\beta\langle S_i^\alpha\rangle
 -\langle S_{i}^\alpha\rangle\langle S_j^\beta\rangle.
\end{eqnarray}
Here the mean-field value of the magnetization on the $i$ site, $\langle S_i\rangle$, is taken as a classical vector
\begin{eqnarray}
\label{Eq:SQ}
  \langle S_i \rangle &=& {\textbf R}\cos({\textbf Q}\cdot r_i)+{{\bf I}}\sin({\textbf Q}\cdot r_i),
\end{eqnarray}
with ${\textbf R}=(R^x,R^y,R^z), {\textbf I}=(I^x,I^y,I^z)$ being the mean-field parameters and ${\textbf Q}=(Q_x,Q_y,Q_z)$ being the magnetic wave-vector.
After performing the mean-field approximation and substituting Eq. (\ref{Eq:SQ}) in the mean-field Hamiltonian,
a discrete Fourier transformation
\begin{eqnarray}
\label{Eq:Fourier}
  C_{k\sigma} &=& \frac{1}{\sqrt{N}}\sum_{i}{\exp{(i\textbf{k}\cdot\textbf{r}_i})}
\end{eqnarray}
is performed, and the mean-field Hamiltonian becomes
\begin{eqnarray}
\label{Eq:HMF}
  H_{MF} &=& \sum_{\textbf{k},\sigma}{H'_t{(\textbf{k})}}+H_J^{MF}.
\end{eqnarray}
Here the hopping Hamiltonian is
\begin{eqnarray}
  H'_t(\textbf{k})&=&[(t'_1\cos k_z-\mu)\sigma_0
  +t'_2(m+2-\cos{k_x}-\cos{k_y})\sigma_z
  +t'_{so}\sin{\bf{k}}\cdot\bm{\sigma}]_{\sigma,\sigma'} C_{\bf{k}\sigma}^+C_{\bf{k}\sigma'} \nonumber \\
&&+[(t'_1\cos (k_z+Q_z)-\mu)\sigma_0
  +t'_2(m+2-\cos{(k_x+Q_x)}-\cos{(k_y+Q_y)})\sigma_z  \nonumber \\
&& +t'_{so}\sin{\bf{(k+Q)}}\cdot\bm{\sigma}]_{\sigma,\sigma'} C_{\bf{k+Q}\sigma}^+C_{\bf{k+Q}\sigma'}, \label{Ht}
\end{eqnarray}
and the magnetic interaction is given by
 \begin{eqnarray}
   H_J^{MF} &=&\chi_1
   C_{\textbf{k+Q}\uparrow}^+C_{\textbf{k}\downarrow}+
   \chi_2C_{\textbf{k+Q}\downarrow}^+C_{\textbf{k}\uparrow}+\chi_3 (C_{\textbf{k+Q}\uparrow}^+C_{\textbf{k}\uparrow}
  - C_{\textbf{k+Q}\downarrow}^+C_{\textbf{k}\downarrow})+H.c.\\ \nonumber
&&+E^0_{MF},
 \end{eqnarray}
where the parameters are
 \begin{eqnarray}
\chi_1&=&\frac{1}{2}\left( A_{1}^+\cos{Q_x}+A_{2}^+\cos{Q_y}
+A_{3}^-\cos{Q_z}+A_{4}^-\sin{Q_z}\right),\\ \nonumber
\chi_2&=&\frac{1}{2}\left(A_{1}^-\cos{Q_x}+A_{2}^-\cos{Q_y}
   +A_{3}^+\cos{Q_z}+A_{4}^+\sin{Q_z}\right),\\ \nonumber
\chi_3&=&\frac{1}{2}\left[A_{5}(\cos{Q_x}+\cos{Q_y})+A_{6}\cos{Q_z}\right],
 \end{eqnarray}
with $A_{1}^{\pm}=\pm J_a(iR^y+I^y)-J_b(R^x-iI^x)$,
$A_{2}^\pm=-J_a(R^x-iI^x)\pm J_b(iR^y+I^y)$,
$A_{3}^{\pm}=J_z^1[(R^x-iI^x\pm(iR^y+I^y)]$,
$A_{4}^{\pm}=J_2[\pm(R^x-iR^y)+(iI^x+I^y)]$,
$A_{5}=J_b(R^z-iI^z)$,
$A_{6}=J_z^2(R^z-iI^z)$.
The constant mean-field energy is
\begin{eqnarray}
  E^0_{MF} &=& \frac{1}{2}[(J_b\cos{Q_x}+J_a\cos{Q_y})\Lambda_x
+(J_a\cos{Q_x}+J_b\cos{Q_y})\Lambda_y\\ \nonumber
&&-J_b(\cos Q_x+\cos Q_y)\Lambda_z]-\frac{1}{2}[J_{z}^1(\Lambda_x+\Lambda_y)
+J_{z}^2\Lambda_z]\cos{Q_z}\\ \nonumber
&&-J_2(R^xI^y-R_yI^x)\sin{Q_z},
\end{eqnarray}
with $\Lambda_{\alpha}=(R^\alpha)^2+(I^\alpha)^2, \alpha=x,y,z$.

The mean-field Hamiltonian can be generally expressed as
\begin{eqnarray}
 H_{MF} = \sum_{\textbf{k},\sigma}  \psi^{\dagger}  ({\bf k}) h_{MF} ({\bf k}) \psi ({\bf k})
\end{eqnarray}
with $h_{MF}$ being a $4 \times 4$ matrix and $\psi ({\bf k}) =(C_{{\bf k}\uparrow}, C_{{\bf k}\downarrow}, C_{\textbf{k+Q}\uparrow}, C_{\textbf{k+Q}\downarrow})^T$. From a given $h_{MF}$, $\langle C_{{\bf k}\sigma}^+C_{{\bf k}\sigma'} \rangle$ and $\langle C_{{\bf k+Q}\sigma}^+C_{{\bf k}\sigma'} \rangle$ are determined.
The effective $\bf R, I$, and $\textbf{Q}$ are then self-consistently with the following self-consistent equations
\begin{eqnarray}
\label{Eq:selfcon-1}
  \frac{1}{N}\sum_{k\sigma}{\langle C_{{\bf k}\sigma}^+C_{{\bf k}\sigma}
+C_{{\bf k+Q}\sigma}^+C_{{\bf k+Q}\sigma} \rangle}=n,
\end{eqnarray}
and
\begin{eqnarray}
\label{Eq:selfcon-2}
  R^\alpha &=& \textrm{Re}(\langle S_Q^\alpha \rangle),
  I^\alpha= \textrm{Im}(\langle S_Q^\alpha \rangle),
\end{eqnarray}
where
\begin{eqnarray}
\label{Eq:selfcon-3}
  S_Q^x &=& \frac{1}{2}\sum_{k}{(C_{\textbf{k}\uparrow}^+
  C_{\textbf{k+Q}\downarrow} +C_{\textbf{k}\downarrow}^+
  C_{\textbf{k+Q}\uparrow})},\\ \nonumber
  S_Q^y &=& \frac{1}{2i}\sum_{k}{(C_{\textbf{k}\uparrow}^+
  C_{\textbf{k+Q}\downarrow}-C_{\textbf{k}\downarrow}^+
  C_{\textbf{k+Q}\uparrow})},\\ \nonumber
  S_Q^z&=& \frac{1}{2}\sum_{k}{(C_{\textbf{k}\uparrow}^+
  C_{\textbf{k+Q}\uparrow}  -C_{\textbf{k}\downarrow}^+
  C_{\textbf{k+Q}\downarrow})}.
\end{eqnarray}
Here $n$ is the density of electron number. By minimizing the free energy, optimal values of
$\bf{R}$, $\bf I$ and $\textbf{Q}$ are finally obtained.

\section{Magnetic phase diagram and topological electronic structures}
\label{Sec:result}
\begin{figure}
  \centering
  \includegraphics[width=5 in]{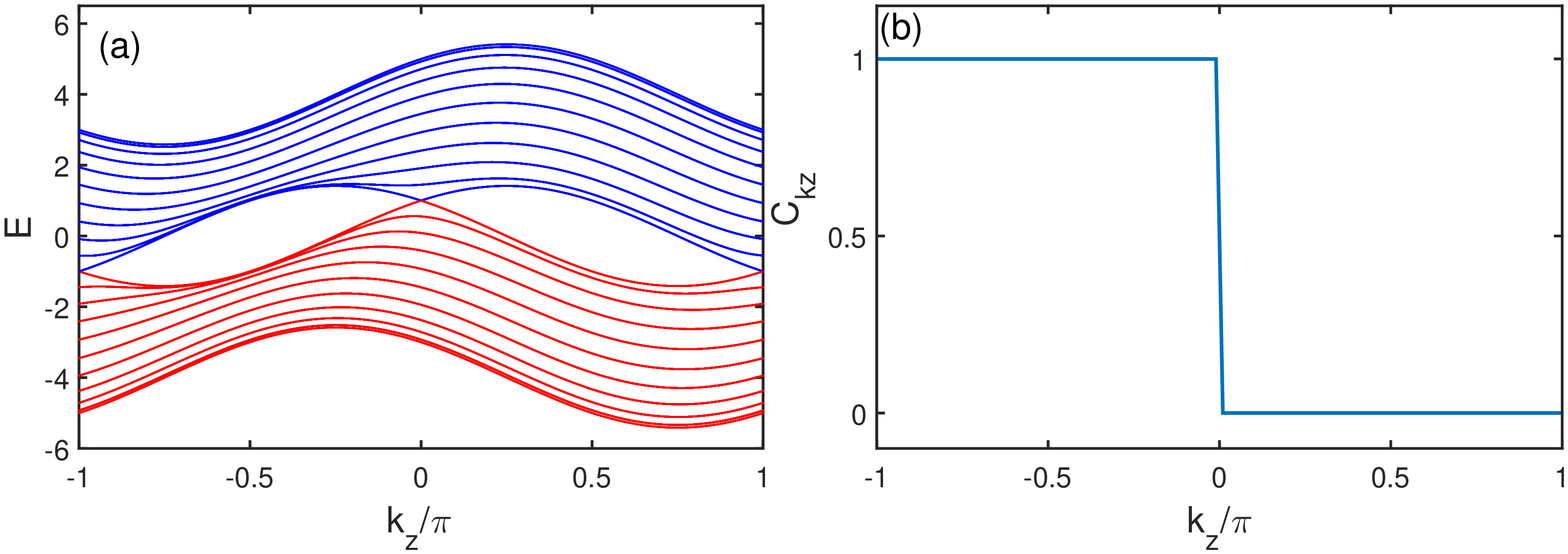}\\
  \caption{(a) Band energies and (b) the Chern number $C_{k_z}$ as a function of $k_z$ with $U=0$, $m=0$,
  $t_1=1$, $t_2=1$ and $t_{so}=1$, and $k_x=k_y$ in (a).}
  \label{Fig:1}
\end{figure}

In this section, we examine magnetic phases that are allowed in the mean-field theory.
For simplicity, we shall set $m=0$. For the case of $U=0$, two linear Weyl nodes are located at
$\textbf{k}=(0,0,0)$ and $\textbf{k}=(0,0,\pi)$.
The chirality of the Weyl node $\textbf{k}=(0,0,0)$ is +1, since the
effective Hamiltonian around which can be
written as $H=t_1+t_{so}\bf{k}\cdot\bf{\sigma}$, while the chirality of $\textbf{k}=(0,0,\pi)$
is $-1$.
The band structure of $U=0$ and $m=0$ is shown in Fig.1 (a).
It is clear that two Weyl points are separated in energy space due to the absence of the
inversion symmetry.
If one fixes $k_z$, the Hamiltonian, $H_{k_z}(k_x,k_y)$,  can be viewed as a 2D system, which is
gapped when $k_z\neq 0,\pi$. In this case, the Chern number $C_{k_z}$ for a fixed $k_z$ is well defined
and can be computed as
\begin{eqnarray}
  C_{k_z} &=& \frac{1}{2\pi}\int_{BZ}{\Omega_{xy}^{(n)}(\bf{k})}dk_xdk_y,
\end{eqnarray}
where the Berry curvature is given by
\begin{eqnarray}
  \Omega_{xy}^{(n)}(\bf{k}) &=& -\textrm{Im}\left[\sum_{n'\neq n}\frac{\langle \psi_{n}(\textbf{k})|\frac{\partial E_k}{\partial k_x}|\psi_{n'}(\textbf{k})\rangle\langle \psi_{n'}(\textbf{k})|\frac{\partial E_k}{\partial k_y}|\psi_{n}(\textbf{k})\rangle
}{[E_{n'}(\textbf{k})-E_n(\textbf{k})]^2}-k_x\leftrightarrow k_y\right].
\end{eqnarray}
Here $E_n(\textbf{k})$ and $\psi_{n}(\textbf{k})$ are the nth eigenenergy and corresponding eigenstate for a given
${\bf k}$ in the Brillouin zone.
Since the Weyl node is a magnetic monopole of the Berry curvature, the
value of $C_{k_z}$ jumps when one goes across Weyl nodes, and the Chern number
of the Chern insulator equals the net monopole charge between the 2D system
defining the trivial and Chern insulator.
As shown in Fig.\ref{Fig:1} (b), $C_{k_z}=0$ when $k_z\in (0,\pi)$,
and $C_{k_z}=1$ when $k_z\in (-\pi,0)$.

\begin{figure}
  \centering
  \includegraphics[width=5.5 in]{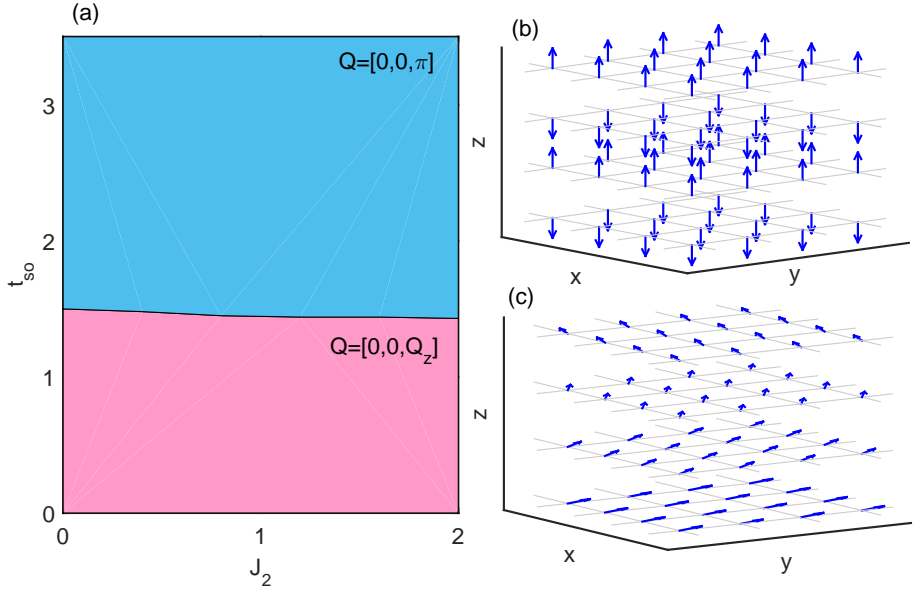}\\
  \caption{(a)Magnetic phases of Weyl semimetal in the strong correlation limit and
the corresponding spin structures of (b) $Q=[0,0,\pi]$ (A-AFM phase )and (c) $Q=[0,0,Q_z]$ (SSDW phase) with $m=0$ and $t_1=t_2=1$.
These magnetic phases are generic with the  A-type AFM phase being expanded if the electron doping density increases.}
  \label{Fig:phase}
\end{figure}
The magnetic phases in the strong correlation limit with $m=0$ is shown in Fig.\ref{Fig:phase} (a).
We find that there are two distinct magnetic phases depending on the
values of $t_{so}$ and $J_2$ in the strong correlation limit.
In the upper blue region, the magnetic wave-vector $Q$ is $[0,0,\pi]$. This
is the A-AFM phase with the magnetic order between the nearest neighbors being FM
in the $xy$ plane and being AFM between layers.
In this phase, the spin orientation is along the $[0,0,1]$ direction,
and the spin structure is shown in panel (b) of Fig.~\ref{Fig:phase}.
Meanwhile, in the below red region, the magnetic wavevector $Q$ is $[0,0,Q_z]$ with
$Q_z$ being a value that is incommensurate with the lattice and is along the $[0,0,1]$ direction.
In this phase, spins are non-collinear and are in the SSDW phase, in which
the spin orientation is in the $xy$ plane and the direction of spiral along $z$ axis as shown in panels (c)
of Fig.~\ref{Fig:phase}.
The A-AFM and SSDW phases are two generic phases. When the electron doping density changes, both phases persist with
one magnetic wavevector being fixed at $Q=[0,0,\pi]$ and the other one being
$Q=[0,0,Q_z]$ with slightly different $Q_z$.
However, as the electron numbers increase, the A-AFM phase expands.

\begin{figure}
  \centering
  \includegraphics[width=4.5 in]{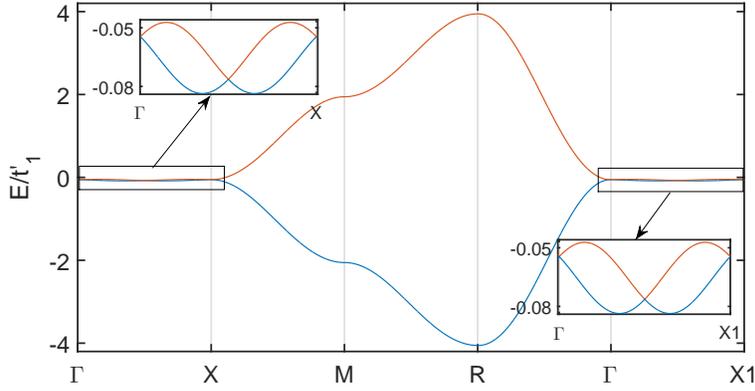}\\
  \caption{Electronic structure of the A-AFM phase with $t_{so}=2$
  and $J_2=1$, displayed along the path: $\Gamma$ $\rightarrow$ X: (0,0,$\pi$)
   $\rightarrow$ M: (0,$\pi$,$\pi$) $\rightarrow$ R: ($\pi,\pi,\pi$) $\rightarrow$
   $\Gamma$ $\rightarrow$ X1: $(0,0,-\pi)$. The electronic structures around the
   four Weyl points $k=(0,0,0),(0,0,\pi)$, and $(0,0,\pm\frac{1}{2}\pi)$ are shown in the inserts.}
  \label{Fig:3}
\end{figure}
\subsection{Topological electronic structures of the A-AFM phase}
We now turn to examine the band structure of the A-AFM phase. Since the
spin orientation is always along the $z$ direction in the A-AFM phase,
the mean-field parameters in $x$ and $y$ direction vanish, $R^x=I^x=R^y=I^y=0$.
In the basis of $\psi^{\dagger}= (C^{\dagger}_{\textbf{k}\uparrow},C^{\dagger}_{\textbf{k}\downarrow},C^{\dagger}_{\textbf{k+Q}\uparrow},
C^{\dagger}_{\textbf{k+Q}\downarrow})$, the magnetic interaction of the Hamiltonian Eq.(\ref{Eq:HMF}) becomes
\begin{eqnarray}
\label{Eq:H00pi}
  H_J^{MF} &=&A(C_{{\bf k}\uparrow}^+C_{{\bf K+Q}\uparrow}
    -C_{{\bf k}\downarrow}^+C_{{\bf K+Q}\downarrow})+H.C. +E^0_{MF},
\end{eqnarray}
where $A=R^z(J_z^2+2J_b)/2$ is independent of $\bf{k}$. In this case,
it is easy to see that two eigenvalues of the Hamiltonian matrix $H_{MF}$
are around the value of $A+E^0_{MF}$ and the other two are around the value of $-A+E^0_{MF}$.
Since $A$ is much greater than $t'_1,t'_2$ and $t'_{so}$ for the A-AFM phase, we find that the system is fully gapped with
the lower two and upper two bands being separated by a gap being around $2A$.

In Fig.\ref{Fig:3}, we illustrate the structure of the lower two bands in the A-AFM phase with $t_{so}=2$ and $J_2=1$.
There are four touching points between the lower two bands, which are
located at $\textbf{k}=[0,0,0], [0,0,\pi]$ and $[0,0,\pm\pi/2]$.
These touching points are still the linear Weyl points with monopole charge being $\pm 1$.
To illustrate this, we plot the Chern number $C_{k_z}$ in Fig.4. It is seen that as $k_z$ changes, whenever energy bands touch, the corresponding Chern number also changes. Effectively, band inversion occurs as $k_z$ changes.
In the region $-\pi <k_z<-\pi/2$ and $0<k_z<\pi/2$, the Hall conductance is quantized with $\sigma_{xy}(k_z) =  e^2/\hbar C_{k_z}$.
Changes of Chern numbers crossing the Weyl points are always $\pm 1$, which demonstrates that the monopole charges of Weyl nodes are $\pm 1$.

In this phase, the magnetic ordering is similar to the proposed model of magnetically doped multilayer heterostructure composed of layers of normal and topological
insulators \cite{Burkov}.
However, different with the proposed model, the magnetic ordering between different layers is AFM
for the A-AFM phase. Since the folding of Brillouin zone in the A-AFM phase, the interaction between
the original Weyl nodes located at $k=(0,0,0)$, and $(0,0,\pi)$ can lead to the new Weyl nodes.

\begin{figure}
  \centering
  \includegraphics[width=4.5 in]{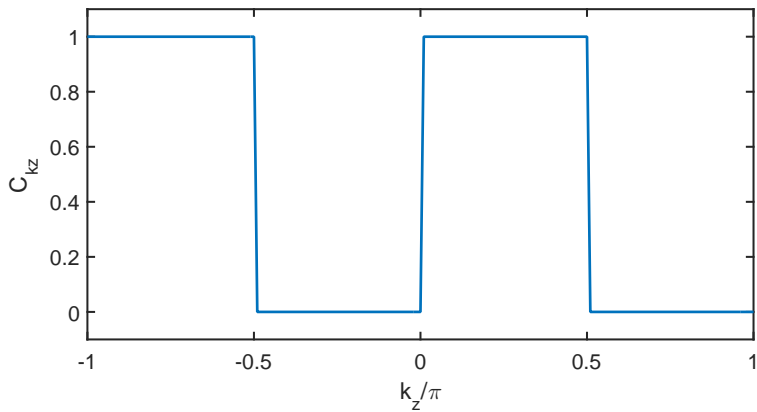}\\
  \caption{Chern number $C_{kz}$ of the A-AFM phase.}\label{Fig:4}
\end{figure}

\subsection{Topological electronic structures of the $[0,0,Q_z]$ phase }
In the SSDW phase, we have the mean field parameters $R^z=I^z=0$ and $R^xI^y+R^yI^x=0$.
As a result, the magnetic interaction of the mean-field Hamiltonian
in the basis of $\Psi^{\dagger}=[C^{\dagger}_{\textbf{k}\uparrow},C^{\dagger}_{\textbf{k}\downarrow},
C^{\dagger}_{\textbf{k+Q}\uparrow},C^{\dagger}_{\textbf{k+Q}\downarrow}]$ becomes
\begin{eqnarray}
\label{Eq:H00spiral}
  H_J^{MF} &=& B C_{{\bf k}\downarrow}^+C_{{\bf{k+Q}}\uparrow}+H.C.+E^0_{MF},
\end{eqnarray}
where $B=[J_{z}^1\cos Q_z-J_2\sin Q_z-(J_a+J_b)]R_x$ is independent of $\bf{k}$.

\begin{figure}
  \centering
  \includegraphics[width=4.5 in]{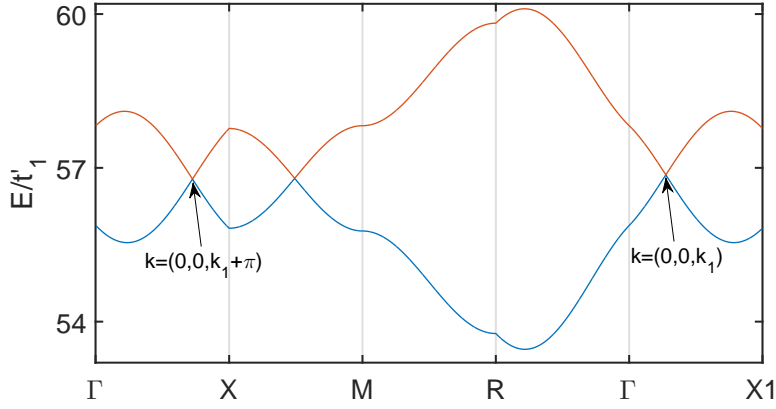}\\
  \caption{Electronic structure of the SSDW phase, displayed along the path: $\Gamma$ $\rightarrow$ X: (0,0,$\pi$) $\rightarrow$ M: (0,$\pi$,$\pi$) $\rightarrow$ R: ($\pi,\pi,\pi$) $\rightarrow$$\Gamma$ $\rightarrow$ X1: $(0,0,-\pi)$. Here, the optimal $Q$ is $[0,0,0.55\pi]$, and $k_1= -0.275\pi$.
  }\label{Fig:5}
\end{figure}

The eigenvalues of the Hamiltonian Eq.(\ref{Eq:H00spiral}) can be approximately derived as
\begin{eqnarray}
E^1_k &\approx & t'_1\cos{k_z}+t'_{so}\sin{k_z}+t'_2(2-\cos{k_x}-\cos{k_y})+E^0_{MF}, \nonumber \\
E^2_k  & \approx & t'_1\cos{(k_z+Q_z)}+t'_{so}\sin{(k_z+Q_z)}+t'_2(2-\cos{k_x}-\cos{k_y})+E^0_{MF}, \nonumber \\
E^3_k  & \approx & B+\frac{1}{2}[t'_1\cos{k_z}+t'_1\cos{(k_z+Q_z)}-t'_{so}\sin{k_z}+t'_{so}\sin{(k_z+Q_z})]+E^0_{MF}, \nonumber \\
E^4_k  & \approx & -B+\frac{1}{2}[t'_1\cos{k_z}+t'_1\cos{(k_z+Q_z)}-t'_{so}\sin{k_z}+t'_{so}\sin{(k_z+Q_z})]+E^0_{MF}. \label{approxE}
\end{eqnarray}
Therefore, in the SSDW phase, two electronic bands of the Weyl semimetal are around $\pm B$ and the middle two bands are around the chemical potential.
Since $B$ is much greater than $t'_1,t'_2$ and $t'_{so}$ in the SSDW phase,
the two energy bands around $\pm B+E^0_{MF}$ are fully gapped, and the band touching points occur
only between the two energy bands around the chemical potential.
In Fig.\ref{Fig:5}, we show the electronic structures of the middle two bands near the chemical potential for
$t_{so}=0.8$ and $J_2=1$.
It is clear that two energy bands touch at two points along $z$ axis with
$\textbf{k}=[0,0,k_1]$ and $\textbf{k}=[0,0,k_1+\pi]$. Here by using Eqs.(\ref{approxE}), the nodal momentum $k_1$ is related to the magnetic wave-vector as
\begin{eqnarray}
k_1=\tan^{-1}\frac{\cos{Q_z}-1}{\sin{Q_z}}.
\end{eqnarray}

\begin{figure}
  \centering
  \includegraphics[width=4.5 in]{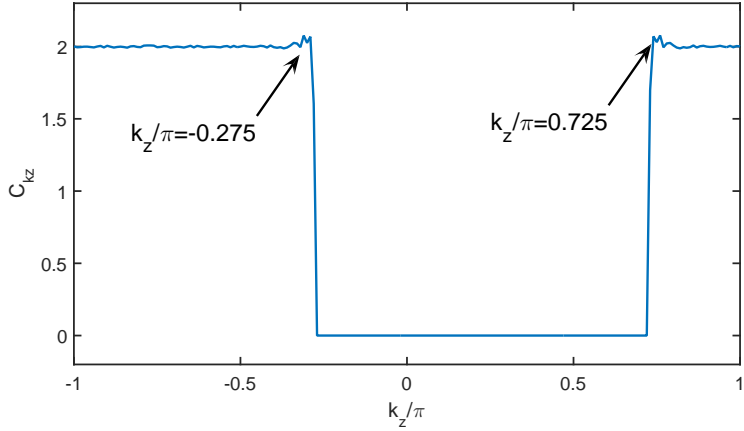}\\
  \caption{Chern number $C_{kz}$ as a function of $k_z$ for the SSDW phase.}\label{Fig:6}
\end{figure}
Around the band touching points, an effective $2\times 2$ Hamiltonian can be found by taking only two
bands near the Fermi energy. For this purpose,
it is convenient to combine the magnetic mean-field Hamiltonian, Eq. \ref{Eq:H00spiral}, with the hopping Hamiltonian, Eq.(\ref{Ht}). In the bias of $\Psi^+=[C^+_{\textbf{k}\uparrow},C^+_{\textbf{k+Q}\downarrow},
C^+_{\textbf{k+Q}\uparrow},C^+_{\textbf{k}\downarrow}]$, around the Weyl node of ${\bf k}=(0,0,k_1)$,
the Hamiltonian matrix, $h_{MF}$, can be written as
\begin{eqnarray}
  h_{MF} &=&\left( \begin{array}{cc}
                  H_1 & V \\
                  V & H_2 \\
                \end{array}
 \right),
  \label{Eq:H00spiral2}
\end{eqnarray}
with
\begin{eqnarray}
  H_1 &=& [t'_1(\cos{k_1}-q_z\sin{k_1})+t'_{so}(\sin{k_1}+q_z\cos{k_1})]\sigma_z^+ , \nonumber \\
  &&+[t'_1(\cos{k_{Q1}}-q_z\sin{k_{Q1}})-t'_{so}(\sin{k_{Q1}}+q_z\cos{k_{Q1}})]\sigma_z^- , \nonumber \\
  H_2&=&[t'_1(\cos{k_{Q1}}-q_z\sin{k_{Q1}})+t'_{so}(\sin{k_{Q1}}+q_z\cos{k_{Q1}})]\sigma_z^+, \nonumber \\
  &&+[t'_1(\cos{k_{1}}-q_z\sin{k_{1}})-t'_{so}(\sin{k_{1}}+q_z\cos{k_{1}})]\sigma_z^-+B \sigma_x, \nonumber \\
  V&=&k_x\sigma_x+k_y\sigma_y. \label{spiralH}
\end{eqnarray}
Here $\sigma_z^{\pm}=\frac{1}{2}(\sigma_0\pm\sigma_z)$, $q_z=k_z-k_1$, and $k_{Q1}=k_1+Q_z$.
In Eq.(\ref{spiralH}), the dominant term is $H_1$. By treating $V$ as the perturbation,  in the second-order perturbation theory,  the effective Hamiltonian around $\textbf{k}=(0,0,k_1)$ is obtained as
\begin{eqnarray}
  H_{eff} &=& H_1-VH_2^{-1}V  \nonumber \\
  &=&-\frac{{t'_{so}}^2}{C}(q_-^2\sigma_++q_+^2\sigma_-)
 +q_z(-t'_1\sin{k_1}+t'_{so}\cos{k_1})\sigma_z  \nonumber \\
 &&+(t'_1\cos{k_1}+t'_{so}\sin{k_1})\sigma_0, \label{effectiveH}
\end{eqnarray}
where $q_{\pm}=k_x\pm ik_y$ and $\sigma_{\pm}=\frac{1}{2}(\sigma_x\pm\sigma_y)$.
Similarly, following the same produce, the effective Hamiltonian around
the other Weyl node ${\bf k}=(0,0,k_1+\pi)$ can be also obtained in the same way.

As pointed in Ref.[\onlinecite{Fang}], the effective Hamiltonian, Eq.(\ref{effectiveH}) describes a
Weyl node carrying monopole charge $-2$, which is protected by the $C_6$ symmetry.
Indeed, as shown in Fig.\ref{Fig:6}, we plot the Chern number $C_{k_z}$. It is seen that as $k_z$ changes, whenever
energy bands touch, the corresponding Chern number also changes. Effectively, band inversion
occurs as $k_z$ changes. In the region $-\pi <k_z<-k_1$ and $k_1<k_z<\pi$, the Hall conductance is quantized with $\sigma_{xy}(k_z) = 2 e^2/h$.
Changes of Chern numbers crossing the Weyl points are always $\pm 2$, which demonstrates that monopole charges of the Weyl nodes are $\pm 2$. Hence the pair of Weyl nodes are double-split Weyl nodes.
In this SSDW phase, the spin ordering can not break the $C_4$ rotation symmetry,
which means such double-Weyl nodes are not protected by the rotation symmetry.
Therefore, as the spin-orbit interaction increases, a magnetic transition occurs with the magnetic order being
turned into the A-AFM order, which breaks the rotational symmetry of the SSDW phase.
As a result, the double-Weyl nodes at $(0,0,\pm k_1)$ are no longer stable\cite{Fang} and are split into four single Weyl nodes, located at $(0,0,0)$, $(0,0, \pm \pi/2)$, and $(0,0,\pi)$.

\section{Discussion and conclusion}
In conclusion, we have explored the magnetic phase and the corresponding topological
electric structures of the Weyl semimetal in the strong onsite Hubbard $U$ limit.
For the minimum model of Weyl semimetal that possesses two linear Weyl nodes, 
the magnetic phases in small $U$ regime can be also analyzed in the mean 
field approximation. In this case, the order parameters at site $i$ are defined as $\langle S_i^+ \rangle= \langle C_{i \uparrow}^{\dagger} C_{i\downarrow} \rangle$, $\langle S_i^- \rangle= \langle C_{i \downarrow}^{\dagger} C_{i\uparrow} \rangle$, and $\langle S_i^z \rangle= \langle (n_{i \uparrow} - n_{i\downarrow})/2 \rangle$. The mean field Hamiltonian is given by
\begin{eqnarray}
  H_M &=& E^0_{HF}+\sum_{{\bf k}, \alpha, \beta} C^{\dagger}_{\alpha} ({\bf k}) H_{0, \alpha \beta} ({\bf k}) C_{\beta} ({\bf k}) +U\sum_{i}{\hat n_{i\uparrow}\hat n_{i\downarrow}}  \nonumber \\
&-&\frac{1}{3} U \sum_{\bf k} \left[ \gamma_1 C_{{\bf k+Q}\downarrow}^+C_{{\bf k}\uparrow}
 + \gamma_2 C_{{\bf k}\downarrow}^+C_{{\bf K+Q}\uparrow} +\gamma_3 C_{{\bf k+Q}\uparrow}^+C_{{\bf k}\uparrow}
 - \gamma_3 C_{{\bf k+Q}\downarrow}^+C_{{\bf K}\downarrow} + H.c.\right].
 \label{model_3}
 \end{eqnarray}
Here $E^0_{MF} =nU/2+ (\Lambda^2_x+\Lambda^2_y+2\Lambda^2_z)/6$, $\gamma_1 =R_x+I_y+i R_y-i I_x$, $\gamma_2 =R_x-I_y+i R_y+i I_x$, $\gamma_3 =R_z-i I_z$, and $H_0$ is given by Eq.(\ref{model_2}) with $R_{\alpha}$ and $I_{\alpha}$ being by Eq.(\ref{Eq:SQ}) and $\Lambda_{\alpha}=(R_\alpha)^2+(I_\alpha)^2$. 
\begin{figure}
  \centering
  \includegraphics[width=4.5 in]{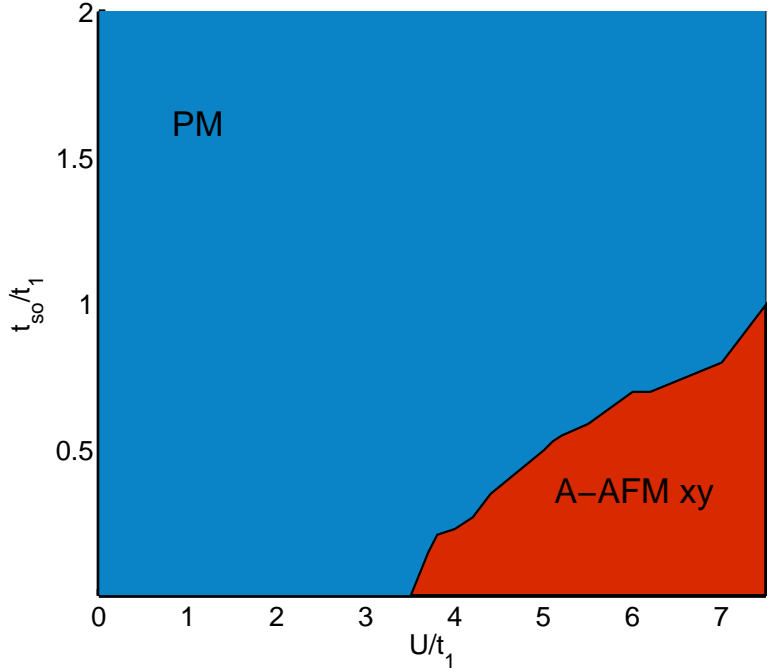}\\
  \caption{Magnetic phases of the Weyl semimetal at half filling ($n=1$) in the weak on-site interaction regime. Here the blue region is the paramagnetic phase. The red region is the A-AFM phase with staggered magnetization being in the $xy$ plane.}
  \label{Fig:phase2}
\end{figure}
Fig.~\ref{Fig:phase2} illustrates possible phase in the weak on-site interaction regime. It is seen that when $U/t_1$ is smaller than $3.5$, the Weyl semimetal is in the paramagnetic (PM) phase. In the PM phase, the topological properties of the band structure are not changed. When $U/t_1$ is larger than $3.5$ and $t_{so}$ is small, the A-AFM phase emerges. However, the A-AFM phase is different from the A-AFM phase in the strong correlation limit. The staggered magnetization lies in the $xy$ plane for the A-AFM phase in the weak interacting regime and is denoted as the A-AFM xy phase. Hence it is clear that when the on-site $U$ is turned on, the Weyl semimetal is paramagnetic without magnetic orders.
The electronic structure is similar to that of the non-interacting Weyl fermions with parameters be just renormalized. As $U$ increases and is strong enough, the Weyl semimetal starts to become magnetic. The increase of $U$ trends to first stabilize the A-AFM xy phase
due to its commensurate nature. As $U$ becomes large, the Dzyaloshinskii-Moriya interaction is induced so that the spiral spin density wave (SSDW) state start to emerge.
Only in the strong on-site $U$ limit, both the SSDW state and A-AFM phase are stabilized.

In the strong $U$ limit, we have derived an extended $t-J$ model. The mean-field phase diagram
in the large-U limit of Hubbard model is established. We show that due to the Dzyaloshinskii-Moriya interaction induced by the spin-orbit interaction,
the A-AFM phase and the SSDW phase are two generic magnetic phases.
As the spin-orbit coupling increases, a quantum phase transition occurs with the SSDW phase being turned into an A-AFM phase.
In addition, it is shown that the topology of the
electronic structure also undergoes changes as the magnetic phase changes.
In the A-AFM phase, number of linear Weyl nodes increases due to the folding of Brillouin zone,
while for the SSDW phase, linear Weyl nodes combine and are turned into double-Weyl nodes carrying
monopole charge $\pm2$.
It should be noted that, while in the small $U$ regime, the Weyl semimetal
can be either in the paramagnetic (PM) phase or the A-AFM phase with spin in the $xy$ plane.
However, these phases cannot leads to a significant change in the topological properties of the Weyl semimetal.
The increase of $U$ trends to stabilize the spiral spin density wave (SSDW) state and A-AFM phase and increase the magnetic energy gain.
The strong magnetization can lead to the significant change in the topological
properties of Weyl semimetal.
Our findings thus pave a new way to build a double-Weyl semimetal
from the Weyl semimetal.

While so far in this work we only consider Weyl semimetals with two linear Weyl nodes located along the $z$ axis, due to the rotational invariance of Hubbard interaction, we expect that the results are
applicable to pairs of Weyl nodes located along other axes.
In general, number of linear Weyl nodes may exceed two. Our results are applicable to
any pair of linear Weyl nodes located along some axis passing through the $\Gamma$ point. Thus,
we expect our findings of the unusual interplay between the topology of electronic structures
and magnetism are applicable to Weyl semimetals with more than two Weyl nodes.

Even though our results are based on the mean-field theory, the Weyl points in the electronic structures have topological origin, which is reflected as their origin from the
band inversion along the $z$ axis. Hence
it is expected that the topological electronic structures of nodes
are robust even if magnetic fluctuations are included. In real materials, the exact electronic structures depends
on detailed crystal symmetries and may show different detailed structures.  Nonetheless, our results offer an important direction to look for in the interplay between magnetic phases and electronic structures and are left for future experimental confirmations.

\begin{acknowledgments}
This work was supported by the Ministry of
Science and Technology (MoST), Taiwan. We also acknowledge support
from TCECM and Academia Sinica Research Program
on Nanoscience and Nanotechnology, Taiwan.
\end{acknowledgments}




\end{CJK*}

\begin{thebibliography}{99}
\bibitem{Hasan} M. Z. Hasan and C. L. Kane, Rev. Mod. Phys. {\bf82}, 3045 (2010).
\bibitem{Hasan1} M. Z. Hasan and J. E. Moore, Ann. Rev. Condens. Matter Phys. {\bf2}, 55 (2011).
\bibitem{Murakami} S. Murakami, New J. Phys. {\bf 9}, 356 (2007); L.  Balents, Physics {\bf 4}, 36 (2011).
\bibitem{Burkov} A. A. Burkov and L. Balents, Phys. Rev. Lett. {\bf107}, 127205 (2011)
\bibitem{Wan} X. Wan, A. M. Turner, A. Vishwanath, and S. Y. Savrasov, Phys. Rev. B {\bf83}, 205101 (2011).
\bibitem{Qi} P. Hosur and X. Qi, C. R. Physique {\bf14}, 857 (2013).
\bibitem{Wang} Z. Wang, Y. Sun, X.-Q. Chen, C. Franchini, G. Xu, H. Weng, X. Dai, and Z. Fang, Phys. Rev. B {\bf85}, 195320 (2012).
\bibitem{Castro Neto} A. H. Castro Neto, F. Guinea, N. M. R. Peres, K. S. Novoselov, and A. K. Geim, Rev. Mod. Phys. {\bf81}, 109 (2009).
\bibitem{Mou} Po-Hao Chou, Liang-Jun Zhai, Chung-Hou Chung, Chung-Yu Mou, and Ting-Kuo Lee, Phys. Rev. Lett. {\bf 116}, 177002 (2016).
\bibitem{Liu} Z. K. Liu, B. Zhou, Y. Zhang, Z. J. Wang, H. M. Weng,
D. Prabhakaran, S.-K. Mo, Z. X. Shen, Z. Fang, X. Dai, Z. Hussain, Y. L. Chen, Science {\bf343}, 864 (2014).
\bibitem{Xiao} D. Xiao, M.-C. Chang, and Q. Niu, Rev. Mod. Phys. {\bf82}, 1959 (2010).
\bibitem{Kobayashi} K. Kobayashi, T. Ohtsuki, K.-I. Imura, and I. F. Herbut, Phys. Rev. Lett. {\bf112}, 016402 (2014).
\bibitem{Goswami}P. Goswami and S. Chakravarty, Phys. Rev. Lett. {\bf{107}}, 196803 (2011).
\bibitem{Huang} Bor-Luen Huang and Chung-Yu Mou, Eur. Phys. Lett. \textbf{88},
68005, (2009); Bor-Luen Huang, Ming-Che Chang, and Chung-Yu Mou, Phys. Rev.
B \textbf{82}, 155462 (2010); Shi-Ting Lee, Shing-Ming Huang and Chung-Yu Mou, J. Phys.: Condens. Matter {\bf 26}, 255502 (2014).
\bibitem{Yang} L. X. Yang, et al., Nat. Phys. {\bf11}, 728 (2015).
\bibitem{Fang} C. Fang, M. J. Gilbert, X. Dai, and B. A. Bernevig, Phys. Rev. Lett. {\bf108}, 266802 (2012).
\bibitem{Xu} G. Xu, H. Weng, Z. Wang, X. Dai, and Z. Fang, Phys. Rev. Lett. {\bf107}, 186806 (2011).
\bibitem{Jian} S.-K. Jian and H. Yao, Phys. Rev. B {\bf92}, 045121 (2015).
\bibitem{Chang2} M.-C. Chang and M.-F. Yang, Phys. Rev. B {\bf92}, 205201 (2015).
\bibitem{Lai} H.-H. Lai, Phys. Rev. B {\bf91}, 235131 (2015).
BaoKai Wang, Nasser Alidoust, Guang Bian, Madhab Neupane, Daniel Sanchez, Hao Zheng, Horng-Tay Jeng,
 Arun Bansil, Titus Neupert, Hsin Lin, M. Zahid Hasan, Proc. Natl. Acad. Sci. 1514581113 (2016).
\bibitem{Yu} S.-L. Yu, X. C. Xie, and J.-X. Li, Phys. Rev. Lett. {\bf107}, 010401 (2011).
\bibitem{Yamaji} Y. Yamaji and M. Imada, Phys. Rev. B {\bf83}, 205122 (2011).
\bibitem{Yoshida} T. Yoshida, S. Fujimoto, and N. Kawakami, Phys. Rev. B {\bf85}, 125113 (2012).
\bibitem{Huang1} Shin-Ming Huang, Shi-Ting Lee and Chung-Yu Mou, Phys. Rev. B {\bf 89}, 195444 (2014).
\bibitem{Sheehy} D. E. Sheehy and J. Schmalian, Phys. Rev. Lett. {\bf 99}, 226803 (2007).
\bibitem{Sekine} A. Sekine and K. Nomura, Phys. Rev. B {\bf90}, 075137 (2014).
\bibitem{Wei} H. Wei, S.-P. Chao, and V. Aji, Phys. Rev. B {\bf89}, 235109 (2014).
\bibitem{Chang} M.-C. Chang and M.-F. Yang, Phys. Rev. B {\bf91},115203 (2015).
\bibitem{Qi2} X.-L. Qi, Y.-S. Wu, and S.-C. Zhang, Phys. Rev. B {\bf74}, 085308 (2006).
\bibitem{Kao} J.-T. Kao, S.-M. Huang, C.-Y. Mou, and C. C. Tsuei, Phys. Rev. B {\bf91},134501 (2015).
\bibitem{Edegger} B. Edegger, V. N. Muthukumar, and C. Gros, Adv. Phys. {\bf56}, 927 (2007).

\end{thebibliography}
\end{document}